\documentstyle[12pt]{article}

\newcommand{\bp}{{\bf p}}

\newcommand{\bx}{{\bf x}}

\begin{document}

\font\ninerm = cmr9

\baselineskip 14pt plus .5pt minus .5pt

\def\footnoterule{\kern-3pt \hrule width \hsize \kern2.6pt}

\hsize=6.0truein
\vsize=9.0truein
\textheight 8.5truein
\textwidth 5.5truein
\voffset=-.4in
\hoffset=-.4in

\pagestyle{empty}
\begin{center}
{\large\bf Replication Error and Time Evolution of a
Self-Replicating System}
\end{center}

\vskip 1.5cm

\begin{center}
{\bf Dongsu Bak\footnote{\ninerm
\hsize=6.0truein email address: dsbak@mach.scu.ac.kr}}\\
{\it Department of Physics,
University of Seoul,
Seoul 130-743, Korea}
\end{center}

\vspace{1.2cm}
\begin{center}
{\bf ABSTRACT}
\end{center}
A self-replicating system where  the elements 
belonging to a solution category can
replicate themselves by copying their own informations, is considered. 
The information carried by each element is defined by an element of
all the n multiple tensor product of a base space that consists
of m different base elements. We assume that in the replication 
the processes of  copying each base information are the same and  
independent from one another and that 
the copying error distribution in each process is characterized 
by a small variation with a quite small mean value. 
Concentrating on the number fluctuation
of the informations in the copying process, we analyze the 
time evolution of the system. We illustrate the change
of averaged number of informations carried by system objects and 
the variation of the number distribution 
as a function of time. 
Especially, it is shown  that the averaged number of information
grows in general after large number of generations.

\bigskip
\bigskip
\begin{center}
\end{center}
\bigskip
\bigskip

\vfill
\space\space 
UOSTP-99-005 
\space\space
 \space\space
\hfill
\eject

\baselineskip=20pt plus 3pt minus 3pt

\pagestyle{plain}
\pagenumbering{arabic}
\setcounter{page}{1}

\pagebreak[3]
\setcounter{equation}{0}
\renewcommand{\theequation}{\arabic{section}.\arabic{equation}}
\section{Introduction}
\nopagebreak
\medskip
\nopagebreak

One of the essential features of the life system is that most of 
its element do possess an ability to replicate themselves by copying
their own informations. The copying mechanisms in each element are mostly
the same but its effectiveness is not perfect due to chemical and  
quantum fluctuations. If it were perfect, 
there should not be any change
of the species of the system 
in its time evolution.
The random fluctuation involved in this copying process is, hence,   
another
important ingredient of the system in understanding its 
time evolution. 

The natural consequences of the replication with random fluctuation
are of great interest since they may explain main  characteristics of 
our life system. Admitting the complication in treating the real 
life system, one may introduce a simplified version of the system
that, however, comprises  the essential features of our life 
system on the aspect of the replication and the 
distribution of the information numbers.

In this note, we shall consider the following simplified version.
The system consists of objects that carry a characteristic 
information defined as an element belonging to  all the n multiple
tensor products of a base space for all the nonnegative integer
n. We shall then define the information number of an object by N
when it belongs to N-multiple tensor product space. The elements in the 
base space are fixed and their total number, m, is finite. 
In the  real 
life system, for example,  
these base elements correspond to the four base acids. Namely, they are adenine, 
guanine, cytosine, and thymine. The information ensemble space is the 
place where the dynamics of interest occurs.  
A solution (moduli) space is defined as a subset of the information 
ensemble space
with a restriction to the objects carrying the information element
that may replicate themselves. We assume that the replication mechanism of
a base element is 
mostly the same and the unit processes copying base information are
independent events. Due to the chemical and quantum mechanical
fluctuations, there might be copying errors that maps one specific
element of the base space to an element in the ensemble space.
Of course, the errors are quite small but still to be a main source 
of the dynamical flow of the system after generations.  
The errors include the change of the original base element
into another base element, omission of the original base element
and mapping the base element into more than a base element.
The probability of mapping the base element to more than two
base units is quite small compared to 
the total probability of errors
and, hence, the mean of the fluctuation  is very close 
to the case of no errors. 
The mean of the errors 
 is more or less inclined to the direction of 
growing of average numbers 
after one copy event because of the asymmetry reflected 
in the fact that there are no counterparts of mapping a base element 
to more 
than two base elements about the symmetry point of the mapping to just
 one element. 

The unit
time between subsequent generations will be set all the same and 
time flow will be measured by the number of generations.
This implies that all the objects in the system at certain time 
have exactly the same numbers of ancestors born after an   
initial time. The next generation will be defined by the set of two
identical descendents from each object of the original system which 
belongs to the solution category. The solution space is determined
as a function of environment and the system itself. The
properties of the solution space are not known mostly 
for realistic life systems. We
shall introduce a minimal assumption
that the density of  solutions over the total ensemble space can be 
defined as a smooth function. This means that
the solutions are densely distributed over the whole 
information configuration space.

Basic features of the time evolution of the system are as follows.
Because of the copying errors, the consequence of 
all the replications of the objects of the original generation 
 belonging to
the solution space will be the next generation  set whose elements
may  or may not belong to the solution space. After a unit time,
the next generation 
again replicate its elements once they belong to the solution space.
Repeating this process from generations to generations, 
the generations will 
flow in the information 
configuration  space quite as the time flies, but the flow will
be restricted to regions of near solution space.
The errors propagate and the initial set diffuses into the 
new arena in the configuration space. What are the properties of the 
dynamics of the system as the number of generations grows?

In this note, we shall concentrate on the information number 
fluctuation of the system objects as a result of the 
 replication. Especially, we will consider
the averaged information number and their variation 
as a function of time. In this way, we will demonstrate
that there is a finite probability of the appearance of
objects carrying
a large number of information compared to 
that of an initial ancestor.

\setcounter{equation}{0}
\renewcommand{\theequation}{\arabic{section}.\arabic{equation}}
\section{Self replicating system and its evolution}
\nopagebreak
\medskip
\nopagebreak

As mentioned in the introduction, the self-replicating system
consists of objects that carry an information
element belonging to the information configuration space.
To define the configuration space, we first introduce a base space
$B$ with a finite number of elements:
\begin{equation} 
\label{e201}
B=\{b_1, b_2, \cdots, b_m \}.
\end{equation} 
The information configuration space is then
the direct sum of all n-multiple tensor product,
\begin{equation} 
\label{e202}
E=\sum^\infty_{n=0}\oplus B^n,
\end{equation} 
where $B^n$ denotes the n multiple tensor product of the base 
space B. For each element of the information configuration space,
we specify its number of information by the order of the 
tensor product where it belongs. As a subset of the ensemble
space, the solution space $S_k$ is composed of the information elements,
whose objects have an ability to replicate themselves from the $k$-th
generation to the $(k+1)$-th generation.
As remarked earlier, we will assume that the solution space
elements are densely distributed over the information ensemble space,
so that one may define the density of the solutions $\rho (n,k)$ 
\begin{equation} 
\label{e203}
\rho(n,k)=Z_\rho {N( S_k\ \cap B^n)\over N( B^n)}.
\end{equation} 
as a 
smooth function of $n$ 
where $N(A)$ is the number of elements in a set $A$ and 
$Z_\rho$ is a normalization constant.
The number fluctuation involved in the replication event of a 
base element takes the  a distribution 
$d(1+\mu_0,\sigma_0)$
with a mean $1+\mu_0$ and a standard deviation 
$\sigma_0$. 
If one denotes the probability of 
$l$ base elements resulted from a unit base  by $p_l$ for a copy,
one may infer that $p_0\sim p_2 =p \,\,\,(p_2-p_0 \ll p)$, 
$p_3 \ll p$ and the contribution of all the higher mode may be ignored
as explained earlier. By an explicit computation,
one finds that $\sigma_0^2\sim 2p$ and $\mu_0 \sim 2p_3 +p_2-p_0 \ll 
\sigma_0^2$.
Here, the variation $\sigma^2_0$
is much smaller than one, 
and we use an approximation
that one is  describing the distribution as if $n$ is a 
continuous
 parameter.  When one replicates an object carrying
$n$ information, the number fluctuation associated with 
this is described by the normal distribution
\begin{equation} 
\label{e204}
\delta n= z(n(1+\mu_0),\sqrt{n}\sigma_0)
\end{equation}
owing to the  independence of all the replication events as well as
the central limit theorem. The informations of 
the system  is also 
given as a subset $A_k$ of the configuration space at the k-th generation.
The 
distribution of the system informations 
over the ensemble space
at the k-th generation is then described by 
\begin{equation} 
\label{e205}
\phi(n,k)=Z_\phi {N( A_k\ \cap B^n)
},
\end{equation} 
where $\phi(n,0)$ represents the initial distribution of the 
informations and $Z_\phi$ denotes a normalization factor.
With help of the distribution function in (\ref{e204}), 
one finds the 
density distribution of the system in the next generation
is determined  by the diffusion process from 
the set $A_k\cap S_k$
at the k-th generation. Namely, all the element in 
$A_k \cap S_k$ are doubled by replications and the  number distribution 
of each descendent is described by (\ref{e204}). 
This is summarized in terms of the density
description by
\begin{equation} 
\label{e206}
\phi(n,k+1)= \int_0^\infty dn'G(n,n')\rho(n',k)\phi(n',k).
\end{equation} 
where $G(n,n')$ is the propagator of the distribution
\begin{equation} 
\label{e207}
G(n,n')= {1\over \sqrt{2\pi n'}\sigma_0} 
{\rm Exp}{\{-{(n-n'-n'\mu_0)^2\over 2n'\sigma_0^2}\}}.
\end{equation}  
In  (\ref{e206}), the factor two by the  doubling as a result 
of replication
 is absorbed into the normalization factor of the density 
function.
By the mathematical induction, 
the density of the system information at arbitrary
generation is obtained from the initial data by
\begin{equation} 
\label{e208}
\phi(n,k)= \int_0^\infty dn'P(n,n';k)\phi(n',0),
\end{equation}
where the propagator is defined as
\begin{equation} 
\label{e209}
P(n,n';k)=\left(\prod_{i=1}^{k-1}  \int_0^\infty dn_i\right)\left(
\prod_{j=0}^{k-1}G(n_{j+1},n_j)\rho(n_j,j)\right),  
\end{equation}
with $n_k=n$ and $n_0=n'$.
The propergator in (\ref{e206}) can be represented in 
a differential form for small $\mu_0$ and $\sigma^2_0$. To measure
the smallness of the mean and the variation, we introduce 
new parameters $\mu$ and $\sigma$ by
\begin{equation} 
\label{e210}
\mu_0=\mu\epsilon, \ \ \ \sigma^2_0=\sigma^2 {\epsilon}
\end{equation}
such that the new parameter $\sigma^2$ is $O(1)$.
To obtain the differential form of the propagation, we define
the infinitesimal time evolution by
\begin{equation} 
\label{e211}
\psi(s,t+\epsilon)= \int_0^\infty ds\, G(s,s')\psi(s',t),
\end{equation} 
where one measures the time by $\epsilon$ multiplied by
the number of generations $k$.
Introducing a variable $\xi$ by
\begin{equation} 
\label{e212}
 {s-s'(1+\mu\epsilon)\over \sigma\sqrt{s'}}=\xi
{\sqrt{\epsilon}} ,
\end{equation} 
one may rewrite the above integral as
\begin{eqnarray} 
\label{e213}
\!\!\!\!\!\!\!\!&&\!\!\!\!\!\!\!\!\psi(s,t+\epsilon)=
\int_{-\infty}^\infty d\xi {(1+b\xi)\over  (1+\mu\epsilon)\sqrt{2\pi}}
e^{-\xi^2/2}
\psi\left({(1+2b\xi +2b^2\xi^2)s\over 1+\mu\epsilon},
t\right) +O(\epsilon^2)
\nonumber\\
\!\!\!\!\!\!\!\!&&\!\!\!\!\!\!\!\!=\!
\int_{-\infty}^\infty \!d\xi e^{-\xi^2/2}{(1+b\xi)\over
\sqrt{2\pi}} 
[(1\!-\!\mu\epsilon)\psi\!+\!s(
2b\xi \!+\!{2b^2\xi^2}\!-\!\mu\epsilon )
 \psi'\!+\!{2b^2\xi^2 s^2}\psi'']
\!+\!O(\epsilon^2)
\end{eqnarray} 
with $b=\sigma\sqrt{\epsilon}/\sqrt{4(1+\mu\epsilon)s}$.
Integrating over the $\xi$-variable, one may obtain 
the following differential equation,
\begin{equation} 
\label{e214}
{\partial\over \partial t}\psi(s,t)=
 \left[ {\sigma^2\over 2}(s{\partial^2\over \partial s^2}+
2{\partial\over \partial s}) -\mu s{\partial\over \partial s}-\mu
\right]\psi(s,t),
\end{equation} 
where the terms of $O(\epsilon^2)$ are ignored. 
Alternatively, the above equation may be presented by
\begin{eqnarray} 
\label{e215}
 {\partial\over \partial t}\psi(r^2,t)=-
H(\bp,\bx)\psi(r^2,t)
\end{eqnarray}
with the Hamiltonian
\begin{eqnarray} 
\label{e216}
 H(\bp,\bx)\equiv {\sigma^2\over 8}\bp^2 + {i\mu\over 2}
\bx\cdot \bp +\mu= 
{\sigma^2\over 8}(\bp+{2i\mu\over \sigma^2}\bx)^2 + {\mu^2\over 2\sigma^2}
\bx^2
\end{eqnarray} 
where the definition $r^2=s=x_1^2+x_2^2+x^2_3+x^2_4$ and 
the four dimensional gradient $\nabla\equiv i\bp$ are introduced.
Although the Hamiltonian is not a Hermitian operator,
the time evolution of the system is well defined. 

Inclusion of the density contribution to the time evolution
is a complicated problem. Some speculations  
on the dynamics with some  generic density function will be relegated to 
the conclusion.
Here, let us consider the case that it is possible 
to write the density function 
in a form
\begin{eqnarray} 
\label{e217}
 \rho(s,t)=\rho_0(t)(1-\epsilon U(s,t) +O(\epsilon^2))
\end{eqnarray} 
where the function $U(s,t)$ satisfies the requirement
\begin{eqnarray} 
\label{e218}
 \epsilon \lim_{L\rightarrow \infty}{1\over L}\int_0^L ds\,|U(s,t)|
\ll 1\,. 
\end{eqnarray}
Note that 
the overall factor $\rho_0$ will be absorbed into the normalization
of the distribution function $\phi$ without changing the probability
amplitude. 
Thus the time evolution of the system function
$\phi(s,t)$ for this case can be easily
identified as
\begin{equation} 
\label{e219}
\phi(r^2,t+\epsilon)=[1-\epsilon H(\bp,\bx)][1-\epsilon U(r^2,t)]
\phi(r^2,t) +O(\epsilon^2),
\end{equation} 
from the combination of (\ref{e206}) (\ref{e214}) and (\ref{e217}).
Hence the differential equation describing the time evolution with 
the density function (\ref{e217}) is
\begin{eqnarray} 
\label{e220}
 {\partial\over \partial t}\phi(r^2,t)=-
[H(\bp,\bx)+U(r^2,t)]\phi(r^2,t).
\end{eqnarray}
As discussed in the introduction, the form of the potential $U(s,t)$
may depend upon
the system function $\phi(s,t)$ because the system itself 
works as an environmental element. However, we won't further
discuss
this possibility
in the following  for the simplicity.

\setcounter{equation}{0}
\renewcommand{\theequation}{\arabic{section}.\arabic{equation}}
\section{Solution and statistics of self-replicating system}
\nopagebreak
\medskip
\nopagebreak

In the preceding section, we discuss the time evolution process
and the differential equation governing the system dynamics.
For the case that one may characterize the system by the
potential $U$ for the solution density contribution, the time 
evolution of the system follows the differential 
equation (\ref{e220}).
Let us first consider the case $U=0$, which 
implies that the density function is constant in its argument.
Namely, the solution space distribution  is uniform
over the 
information configuration 
space.
To find the dynamics, it is convenient to construct the kernel
$K(r^2, {r'}^2;t)$
that is defined as a solution of the equation (\ref{e220})
with the  initial condition
\begin{eqnarray} 
\label{e301}
\phi_0(r^2)=\delta(r^2-{r'}^2).
\end{eqnarray}
With help of the kernel, the solution for a general initial condition
$\phi(r^2,0)$ is obtained with
\begin{eqnarray} 
\label{e302}
\phi(r^2,t)=\int_0^\infty d({r'}^2)K(r^2,{r'}^2;t)
 \phi({r'}^2,0).
\end{eqnarray}
In order to find the kernel, we first introduce the
function $\tilde\phi (r^2,t)$ by
\begin{eqnarray} 
\label{e303}
\phi(r^2,t)=e^{{\mu r^2/\sigma^2}}
\tilde\phi({r}^2,t),
\end{eqnarray}
and insert this into the equation (\ref{e220}). One finds that 
the equation for 
$\tilde\phi$ now reads
\begin{eqnarray} 
\label{e304}
 {\partial\over \partial t}\tilde\phi(r^2,t)=-
\tilde{H}(\bp,\bx)\tilde\phi(r^2,t)
\end{eqnarray}
with a new Hamiltonian
\begin{eqnarray} 
\label{e305}
 \tilde{H}(\bp,\bx)= 
{\sigma^2\over 8}\bp^2 + {\mu^2\over 2\sigma^2}
\bx^2.
\end{eqnarray} 
It is interesting to note that this is the Hamiltonian
that describes a simple harmonic oscillator in four
dimensional flat Euclidean space. 
The  kernel for the equation (\ref{e304}) is simply a product of
the propagator for 
one dimensional simple harmonic oscillator\cite{hibbs}:
\begin{eqnarray} 
\label{e306}
\tilde{ K}(\bx,\bx';t)=\left({\mu\over 2\pi \sinh \mu t}\right)^2
{\rm Exp}\left\{ -{2\mu\over \sigma^2 \sinh \mu t} 
[(\bx^2+{\bx'}^2)\cosh\mu t -2\bx\cdot \bx']\right\},
\end{eqnarray} 
with the solution of the differential equation (\ref{e304})
\begin{eqnarray} 
\label{e307}
g(\bx,t)=\int d^4 x'  \tilde{K}(\bx,\bx';t)
 g(\bx',0).
\end{eqnarray}
for an arbitrary initial condition $g(\bx,0)$.
The kernel for our system is then obtained by taking $g(\bx',0)
=e^{-{\mu {r'}^2/\sigma^2}}
\delta({r'}^2-{r_0}^2)$ in (\ref{e307}) and multiplying  
the factor  $e^{{\mu r^2/\sigma^2}}$. Using the integral representation
of the Bessel function\cite{ryzhik}, one finds 
that the expression for the kernel
reads
\begin{eqnarray} 
\label{e308}
&& K(r^2,r_0^2;t)=e^{{\mu r^2/\sigma^2}}
\int d^4 x'  \tilde{K}(\bx,\bx';t) e^{-{\mu {r'}^2/\sigma^2}}
\delta({r'}^2-{r_0}^2)\nonumber\\
&&=
{\mu r_0\over 2 r \sinh \mu t}
{\rm Exp}\left\{ {2\mu\over \sigma^2} 
[(r^2-r_0^2)-(r^2+r_0^2)\coth\mu t]\right\} 
I_1({2\mu r r_0\over \sigma^2 \sinh \mu t}).
\end{eqnarray} 
where $I_1(x)$ is the first-kind  Bessel function of 
imaginary argument.
For the consistency of the equation (\ref{e302}), the kernel 
must satisfy the relation
\begin{eqnarray} 
\label{e309}
K(r^2,{r'}^2;t+t')=\int_0^\infty d({z}^2)K(r^2,{z}^2;t)
K(z^2,{r'}^2;t'),
\end{eqnarray}
which may obtained by applying the equation (\ref{e302}) twice
for time intervals $[0,t']$ and $[t',t+t']$.
Upon using the expression (\ref{e308}),
 this composition rule may be checked
explicitly by a straightforward computation with help of
 the formulas
for the definite integrals involving the Bessel 
function\cite{ryzhik}.

Following the standard rule, the mean $M(t)$
and $V(t)$ variation of the number of information
are defined by
\begin{eqnarray} 
\label{e310}
M(t)&\equiv& {\int_0^\infty ds\, s\, \phi(s,t)\over 
\int_0^\infty ds \, \phi(s,t)}\\
\label{e311}
V(t) &\equiv& {\int_0^\infty ds\, s^2 \,\phi(s,t)\over 
\int_0^\infty ds \, \phi(s,t)} -M^2(t).
\end{eqnarray}
For the case of  the initial condition $\phi_0(s)=\delta(s-s_0)$, 
one may compute explicitly the mean and variation noting
that $\phi(s,t)$ is given by $K(s,s_0;t)$. For this initial condition, 
we first compute
the total amplitude $P(t)$; the resulting expression reads
\begin{eqnarray} 
\label{e312}
P(t) \equiv \int_0^\infty ds \, \phi(s,t)
=
1-{\rm Exp}\{-{2\mu s_0 e^{\mu t}\over \sigma^2 \sinh \mu t}\}
\end{eqnarray}
As a function of $t$, $\mu$
and $\sigma$, the mean and variation for the initial condition are
explicitly
\begin{eqnarray} 
\label{e313}
M(t) &=& {s_0e^{2\mu t}} P^{-1}(t) 
\\
\label{e314}
V(t) &=& \left[ {\sigma^2 s_0 P(t)}e^{3\mu t}\mu^{-1} \sinh \mu t-
s_0^2 {\rm Exp}\{4\mu t -{2\mu s_0 e^{\mu t}\over \sigma^2 \sinh \mu t}\}
\right]
 P^{-2}(t) .
\end{eqnarray}
By taking the limit that $\mu$ goes to zero from 
these expressions, one may obtain  the mean and variation 
of the system with
$\mu=0$ and they are
\begin{eqnarray} 
\label{e315}
M_0(t) &=& {s_0\over 1- e^{-{\lambda \over t}} 
}\\
\label{e316}
V_0(t) &=& {s^2_0e^{\lambda \over t}\over ( e^{\lambda \over t}-1)^2} 
\left[{2t\over \lambda}(e^{\lambda \over t}-1) -1\right].
\end{eqnarray}
where $\lambda=2 s_0/\sigma^2$.
For $t \ll \lambda$, one finds that the mean and variation are  
\begin{eqnarray} 
\label{e317}
M_0(t)\sim s_0,\ \ \ V_0(t)\sim 2s_0^2 t/\lambda,
\end{eqnarray}
which agree with those computed from the initial condition 
$\phi(s,0)=\delta(s-s_0)$. On the other hand, for $t\gg \lambda$,
we have
\begin{eqnarray} 
\label{e318}
M_0(t)\sim {\sigma^2 t\over 2},\ \ \ V_0(t)\sim 
\left({\sigma^2 t\over 2}\right)^2.
\end{eqnarray}
Hence the mean and variation are independent of the initial parameter
$s_0$ and grow in a power law of time  as the time gets larger.
With the nonvanishing $\mu$, one finds that as far as $|\mu t| \ll 1$,
the mean and the variation behave in the same way as the case with 
$\mu=0$. On the other hand, for the case  $\mu t \gg 1$, the mean 
and variation  are approximated by 
\begin{eqnarray} 
\label{e319}
M(t) &\sim& {s_0 e^{2\mu (t +\lambda)}\over e^{2\mu\lambda }-1 
}\\
\label{e320}
V(t) &\sim& {s^2_0e^{2\mu (2t +\lambda)}\over ( 
e^{2\mu\lambda }-1)^2} 
\left[{1\over \mu \lambda}(e^{2\mu\lambda }-1) -1\right].
\end{eqnarray}
For the negative $\mu$, the large time (i.e. $t \gg -\mu^{-1}$ ) behaviors
are simply  $M(t) \sim \sigma^2/(4|\mu|) $ and $V(t)\sim 
\sigma^4/(4\mu)^2$ 
and hence only in this case the mean and variation of the system are 
bounded from above all the time. 

One may further consider the case with nonvanishing potential $U(s,t)$.
Of course, it is not possible to solve the system explicitly without
an explicit form of the potential. However, the influence of  
 the potential to the time evolution of the system is
 not so complicated to understand. 
Since the role of the potential $U$ appears as a weighting factor 
$e^{-\epsilon U(s,t)}$ on the amplitude in each propagating interval
$[t,t+\epsilon]$, the larger positive  value of 
$U(s,t)$ at a certain position 
results in the more suppression of the amplitude, while the larger
negative  
value leads to the more amplification of the probability amplitude. 
Especially, for the case that the harmonic potential term in 
(\ref{e305}) dominates the potential $U$ at large $s$, e.g.
\begin{eqnarray} 
\label{e321}
|U(s,t)|\ll {\mu^2 s/2\sigma^2} \ \ \  {\rm as} \ s\rightarrow 
{\rm large},
\end{eqnarray}  
the large time features of the system  mostly agrees with those in 
(\ref{e319}) and (\ref{e320}). For the case that $U$ is much larger than 
$U_0= {\mu^2 s/2\sigma^2}$, the growth of the mean and variation 
are relatively suppressed. The other limiting case, $-U/ U_0 \gg 1$,
the mean and variation  grow much faster in time compared to the case 
$U=0$.

What happens to the case that the influence of the density 
cannot be presented
in terms of the potential description? In this case, the amplification
by the density multiplication in each generation are $O(1)$ and become
huge after a large number of  generations are passed. The distribution 
function of the system
will be sharply peaked at the maximum 
of the density  after a large number of generations. This is characterized 
by the mean $M(t)\rightarrow {\rm Max} (\rho(s,t))$ with 
$V(t) \rightarrow  0$.  Thus, in this case one may conclude that there 
is no evolution of the system at all after a large time as far as the
probability distribution of the system is concerned. 

One may speculate that the real life system presides in between the two 
limiting cases of  the $O(1)$ variation of the density function and 
$U=0$.

\setcounter{equation}{0}
\renewcommand{\theequation}{\arabic{section}.\arabic{equation}}
\section{Outlook}
\nopagebreak
\medskip
\nopagebreak

As a simplified version of the real life system, we have modeled
the self-replicating system whose dynamics can be projected onto the
space of information configurations. Each object in the system 
carries an information element in the configuration space. The unit copying 
process for the replication of the base information is characterized
by  the number fluctuation with the mean $\mu_0$ and 
the standard deviation 
$\sigma_0$, and each of unit copying is assumed to be an independent
event. We define the solution space by the condition that its element
has an  enough information to replicate itself to the next 
generation. 
To characterize the properties of the solution space, we define the
density function of the solution space over the information 
configuration space.
We have shown that the dynamics of the system
is described with help of  the time evolution kernel and may be mapped
into
the evolution of the four dimensional harmonic oscillator for 
the case of uniform density of the solution space.
In addition, we have discussed the system dynamics  for the various case
of  
nonuniform density.
Especially, we have proved that the time evolution of the mean and 
variation of 
the information numbers over the system configuration are growing 
with time
with a few exceptions.  This exception compiles the case of the 
negative
$\mu$, 
which is improbable in the realistic system.

In our model, 
some details of the real life system are not included for simplicity.
For example, the unit time intervals between generations
are not all the same but depend upon the generations
and the information number variable
in the real life system. The number of 
descendents from a system object after a generation is also dependent on 
the time and the number variable. Moreover, there are two kinds of the 
replication processes, which are nothing but asexual and sexual 
reproductions. There may be also a probability to fail
to produce something else rather than  base elements 
in the unit copying process, though it is expected to be extremely small.
As mentioned earlier, it may be the case that the density of the solution 
space is a functional of the system distribution $\phi(s,t)$. In this 
case, the equation governing the time evolution inevitably becomes
nonlinear in $\phi(s,t)$ and this nonlinear effect may be important 
in understanding the decelerating force of the population growth
due to the limitation of resources. 

 These kind of fluctuations may be important in understanding 
the local
dynamical evolution at a certain local time interval $[t, t+\Delta t]$.
However, if the fluctuations are averaged over the long time, 
the effects of these deviations may be effectively
ignored without introducing any serious change in global pictures. 
Furthermore, the detailed description of these variations may be
incorporated by the slight modifications of the model presented.

Nevertheless, the 
effect of these variation may be crucial in comparing the theory
to the characteristics of the real life system because
observations on the real life system is confined to a present short 
time interval. 
In this sense,
the experimental implication of our model need more investigations
because how to extract essential features of the real system dynamics
from the present data depends considerably on the detailed form of local
fluctuations.

\bigskip
\begin{center}
{\bf ACKNOWLEDGEMENTS}
\end{center}
\ \indent
The authors would like to  thank Choonkyu Lee, Joohan Lee, 
Hyunsoo Min, 
and Jae Hyung Yee for 
enlightening discussions.
\hfill
   
\bigskip
\bigskip



\end{document}